\documentclass[useAMS,usenatbib]{mn2e}

\usepackage{graphics}
\usepackage{graphicx}
\usepackage{amssymb}
\usepackage{bm}
\usepackage{dcolumn}

\def\msunh{\mbox{$h^{-1}M_\odot$}}
\def\msun{\mbox{$M_\odot$}}
\def\mpc{\mbox{$h^{-1}Mpc$}}
\def\beq{ \begin{equation}}
\def\eeq{\end{equation} }
\def\bea{\begin{eqnarray}}
\def\eea{\end{eqnarray}}

\begin{document}

\title[Flattening of c(M) and implications for boost]{The flattening of the concentration-mass relation towards low halo masses and its implications for the annihilation signal boost}
\author[S\'anchez-Conde \& Prada]{Miguel A. S\'anchez-Conde$^{1}$\thanks{E-mail: masc@stanford.edu; f.prada@csic.es} and Francisco Prada$^{2,3,4}$\\
$^{1}$ W. W. Hansen Experimental Physics Laboratory, Kavli Institute for Particle Astrophysics and Cosmology, \\
\hspace{0.5cm}Department of Physics and SLAC National Accelerator Laboratory, Stanford University, Stanford, CA 94305, USA\\
$^{2}$ Campus of International Excellence UAM+CSIC, Cantoblanco, E-28049 Madrid, Spain\\
$^{3}$ Instituto de F\'isica Te\'orica (UAM/CSIC), Universidad Aut\'onoma de Madrid, Cantoblanco, E-28049 Madrid, Spain\\
$^{4}$ Instituto de Astrof\'isica de Andaluc\'ia (IAA-CSIC), Glorieta de la Astronom\'ia, E-18008 Granada, Spain}

\date{}

\pagerange{\pageref{firstpage}--\pageref{lastpage}} \pubyear{2002}

\maketitle

\label{firstpage}

\begin{abstract}
In the standard cold dark matter (CDM) theory for understanding the formation of structure in the Universe, there exists a tight connection between the properties of dark matter (DM) haloes, and their formation epochs. Such relation can be expressed in terms of a single key parameter, namely the halo concentration. In this work, we examine the median concentration-mass relation, c(M), at present time, over more than 20 orders of magnitude in halo mass, i.e. from tiny Earth-mass microhalos up to galaxy clusters. The c(M) model proposed by Prada et al. (2012), which links the halo concentration with the {\it r.m.s.}  amplitude of matter linear fluctuations, describes remarkably well all the available N-body simulation data down to $\sim$10$^{-6} \msunh$ Êmicrohalos. A clear fattening of the halo concentration-mass relation towards smaller masses is observed, that excludes the commonly adopted power-law c(M) models, and stands as a natural prediction for the CDM paradigm. We provide a parametrization for the c(M) relation that works accurately for all halo masses. This feature in the c(M) relation at low masses has decisive consequences e.g. for $\gamma$-ray DM searches, as it implies more modest boosts of the DM annihilation flux due to substructure, i.e., $\sim35$ for galaxy clusters and $\sim15$ for galaxies like our own, as compared to those huge values adopted in the literature that rely on such power-law c(M) extrapolations. We provide a parametrization of the boosts that can be safely used for dwarfs to galaxy cluster-size halos.

\end{abstract}



\section{Introduction}


In the current structure formation scenario, cold dark matter (CDM) perturbations in the primordial density field constitute the seeds of all the structures that we observe in the Universe today. This primordial density field is fully characterized by its power spectrum, $P(k)$, which exact properties were set in the early inflationary stages \citep{inflation1,inflation2} and processes before the recombination epoch \citep{bond84,bbks86}. In the standard $\Lambda$CDM cosmological model \citep{planck13}, the {\it r.m.s.} amplitude of linear fluctuations in the density field, smoothed over a scale R,  $\sigma(R)$, decreases very slowly with mass at small scales (in log space), while it becomes a power-law at larger scales. Since the first objects to collapse are those with the greater {\it r.m.s.} amplitudes, the primordial CDM perturbations naturally lead to a bottom-up structure formation scenario in which the smallest objects form first, while the larger structures collapse later on and arise from merging and accretion of the smaller ones. This hierarchical model of structure formation agrees very well with observations and represents indeed a big success of the $\Lambda$CDM cosmology paradigm (e.g.~\citet{LSS}).\footnote{Yet, the full hierarchical scenario of CDM remains to be tested at the smallest galactic scales (dwarfs). In this regard, allowed Warm Dark Matter (WDM) models with a filtering mass near this scale would predict a different formation scenario for dwarfs \citep{zavala09}.}

The formation of dark matter (DM) halos is a key aspect of the CDM structure formation scenario described above, where the halo formation epoch fully reflects the adopted underlying cosmological model. More importantly, there exists a strong correlation between the inner structural properties of DM halos and their formation epochs \citep{bullock01,wechler02,maccio08,zhao09,ludlow13}. 
The internal halo properties are often described in terms of the $concentration$ parameter, $c \equiv R_{vir}/r_s$, where $R_{vir}$ is the halo virial radius, defined as the one that contains an enclosed density $\Delta_{crit}$ times the {\it critical} density of the Universe (set to $\Delta_{crit}=200$ throughout this work), and $r_s$ is a scale radius \citep{nfw96}. 
A general trend found in $N$-body cosmological simulations is that halos with larger mass exhibit smaller concentration. Furthermore, for the same halo mass,  concentration declines with redshift, given the decrease of the mean matter density of the Universe over time. Yet, despite to all efforts in understanding the concentration-mass relation, c(M), numerical simulations have only explored in detail the most massive tail of the entire halo mass range expected in the CDM scenario, that could be as low as 10$^{-6}~\msun$ or even less, with its exact value being set by the kinetic-decoupling temperature and the mass of the DM particles \citep{green04,minmass}. 
Remarkably, new challenging simulations have been recently performed which provide further insights into the c(M) relation at the smallest scales, i.e. the first structures formed in the Universe \citep{ishiyama10,anderhalden13,ishiyama14}. The works by \citet{ishiyama10} and \citet{anderhalden13}, although outstanding, did not allow them to extract firm conclusions given the poor halo statistics, often being simulated only few of those tiny halos. Only very recently, the work by \citet{ishiyama14} has been able to measure mean microhalo properties using dozens to thousands of them in the mass range between $4\times10^{-4} \msun$ down to $\sim 2\times10^{-6} \msun$. 

In this work, we review our current knowledge of the median c(M) relation, drawn from state-of-the-art N-body cosmological simulations, for the entire halo mass range. 
This exercise will turn out to be extremely useful to highlight where we stand on understanding the nature of halo concentrations and to also identifying new opportunities for further theoretical and numerical investigation. 
We perform such a study at the present epoch, which is particularly useful e.g. for a direct comparison with observations, and also allows for a one-to-one comparison at the same epoch for all the reported results available in the literature. Although data are scarce (or non-existent) at small masses, namely below $\sim$10$^9-$10$^{10}\msunh$, a comparison between both, cosmological simulations and physically motivated c(M) models have allowed us to extract meaningful conclusions. As we will show here, only realistic models that link halo concentration with the amplitude of the linear density field fluctuations $\sigma(M)$, such as the toy model by \citet{bullock01} (or its refined version by \citet{maccio08}), and the most recent by \citet{prada12} (hereafter B01, M08 and P12, respectively), represent a good description of what is measured in simulations at all redshifts. c(M) power-law extrapolations to lower masses, as often found in the literature (e.g., \citet{springelnature,gao11}), are strongly disfavored and lead to wrong conclusions in different contexts. As an example, we will discuss on the implications for the computation of the DM annihilation substructure boosts, which are particularly sensitive to the choice of the c(M) model, and will show that moderate substructure boosts are indeed expected, contrary to what has been recently reported in the literature. 

We organized this work as follows. 
Section \ref{sec:cm0} is devoted to discussing the current knowledge of halo concentrations at present time as given by $N$-body cosmological simulations and comparing the simulation data against the P12 c(M) model. The implications of our studies for gamma-ray DM searches, namely the expected enhancements to the DM annihilation signal due to halo substructure, are detailed in Section \ref{sec:boosts}. Finally, we summarize the main results of our work in Section \ref{sec:summary}.


\section{Halo concentration at the present epoch}  \label{sec:cm0}

\begin{figure*}
\includegraphics*[width=0.93\textwidth]{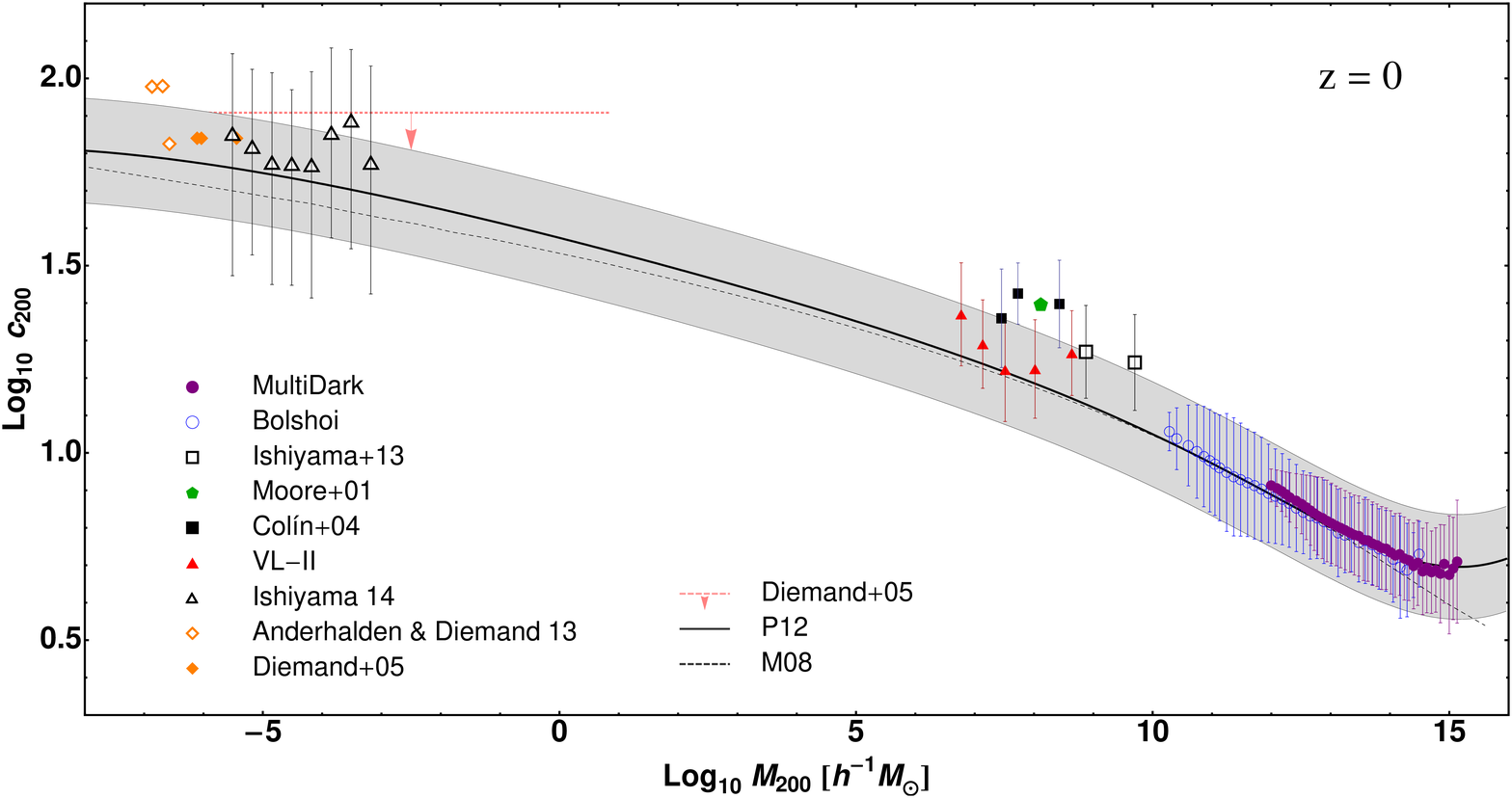}\\
\includegraphics*[width=0.93\textwidth]{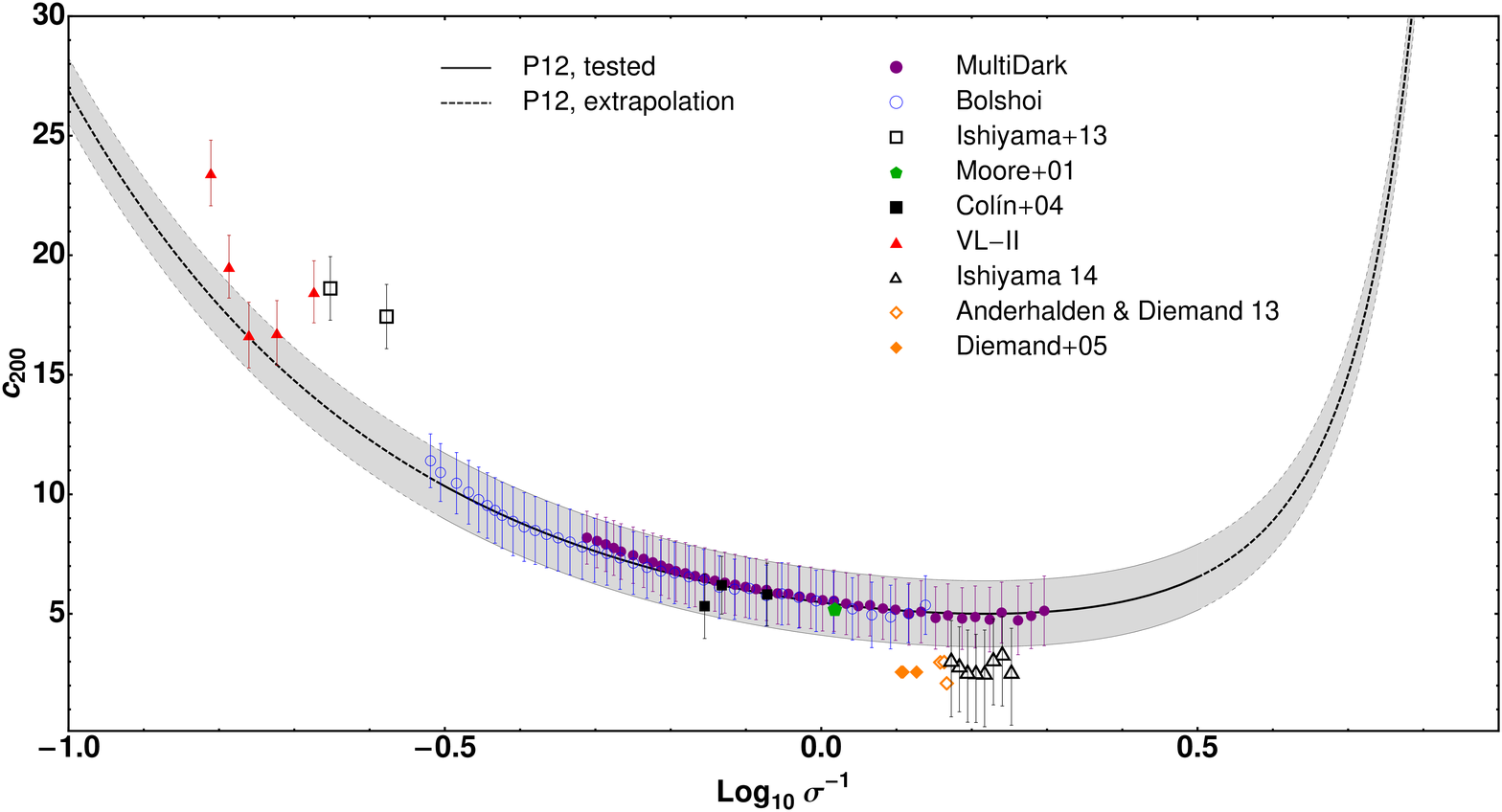}
\caption{{\it Top panel:} Current knowledge of the median concentration-mass relation at $z=0$ for all halo masses available in the literature from different simulation data sets, i.e. from the smallest Earth-like DM microhalos predicted to exist in the CDM universe ($\sim$10$^{-6} \msunh$), up to the largest cluster-size halos ($\sim$10$^{15} \msunh$). At the high-mass end, the results from Bolshoi (blue circles) and MultiDark (purple circles) are shown. The two empty black squares at $\sim$10$^{9} \msunh$ and the three filled black squares at $\sim$10$^{8} \msunh$ were derived from \citet{ishiyama13} and \citet{colin}, respectively. Another individual "Draco««-like 10$^{8} \msunh$ halo is also plotted as a green pentagon \citep{moore01}. A couple hundreds dwarf halos with masses $\sim$10$^{6}$ -- 10$^{9}~\msunh$ (red triangles) were extracted from the VL-II data \citep{VLII}. At the low-mass end, we show the microhalo results taken from \citet{diemand05} (orange filled diamonds) and \citet{anderhalden13}  (orange empty diamonds) for individual halos, as well as those recently reported by \citet{ishiyama14} for a sample of thousands of microhalos (empty black triangles). We also provide the upper limit to halo concentrations obtained by \citet{diemand05} in the range 10$^{-6}$ -- 10$~\msunh$ (pink dotted line). The P12 concentration model \citep{prada12} is shown with a solid line. The shaded grey region represents a typical $1\sigma$ concentration scatter of 0.14 dex centered on the P12 model. The dashed curve represents the updated M08 version \citep{maccio08} of the B01 toy concentration model \citep{bullock01}. All concentration values but those from MultiDark, Bolshoi and VL-II, have been extrapolated down to $z=0$ by means of the $(1+z)$ correction factor. {\it Bottom panel:} Same data set but displayed in the $c$~--~$\sigma^{-1}$ plane, which allows for a more detailed analysis and comparison between simulations and model in terms of the amplitude of linear density fluctuations. The concentration values shown are those in the original set of simulations at the corresponding redshift where they were measured, while the $\sigma(M)$ values are the ones that halos would have at present time for {\it those} values of the concentration, see text for further details. Solid (dashed) line refers to the $\sigma(M)$ range in which the P12 model was (not) tested against simulations.}
\label{fig:cm0}
\end{figure*}

The top panel of Fig.~\ref{fig:cm0} summarizes all that we currently know from $N$-body cosmological simulations about the c(M) median relation, at the present epoch, for all explored halo masses, i.e. from the smallest structures predicted to exist in the CDM universe ($\sim$10$^{-6}\msunh$) up to the largest masses ($\sim$10$^{15}\msunh$). This represents an update of that shown e.g. in Fig.~1 of \citet{colafrancesco06}. Different data sets are drawn from different simulations. Usually, every simulation work was optimized to study DM halo properties at a particular scale or mass range. The data reported in the literature are abundant in the high-mass end (10$^{10}$ -- 10$^{15}\msunh$), 
where mass and force resolution limitations are less severe.\footnote{See e.g., \citet{klypin13} for a complete study on accuracy and conditions for numerical convergence.} Among all the available simulations that study the large scale structure, we show in the top panel of Fig.~\ref{fig:cm0} the median concentration-mass relation and $1\sigma$ errors (halo-to-halo variations)
 found in the Bolshoi \citep{bolshoi} and Multidark \citep{prada12} simulation boxes at $z=0$ (blue and purple circles, respectively). Both Bolshoi and Multidark represent the state-of-the-art $N$-body simulations for the WMAP7 cosmology \citep{wmap7}. Together, they cover the halo mass range between $\log{M (\msunh)}\sim$ 10~--~15. We note that a rather similar coverage in halo mass could also be obtained from the Millennium set of simulations \citep{millennium,millenniumII,millenniumXXL}, although in this case previous WMAP1 cosmological parameters were assumed. Only recently, the work by \citet{ishiyama13}, which adopts the WMAP7 cosmology, has made possible to go down to $\sim5\times10^8 \msunh$ with superb halo statistics. After having corrected by the different halo mass definition, we included in the top panel of Fig.~\ref{fig:cm0} their median value concentrations at $z=0$ plus $1\sigma$-error bars between $\sim 5\times10^8 - 10^{10} \msunh$ (empty black squares), noting that each data point refers to tens of thousands of halos. However, below these halo masses, the simulation data available in the literature is very scarce. Yet, we have made an exhaustive search in our attempt to 
 compile all c(M) data over the wider mass range available for a similar cosmology. A good example of such effort is the three data points with error bars at $\sim$10$^8\msunh$ (filled black squares). These correspond to median concentration values (and their corresponding $1\sigma$ errors) 
 derived from the study performed by \citet{colin} for well resolved dwarf galaxies at $z=3$. We grouped the $\sim50$ objects originally presented in their work in only 3 mass bins, and later extrapolated the concentration values down to $z=0$ multiplying by the factor $(1+z)$ that accounts for the expansion of the Universe, as suggested by, e.g., \cite{bullock01}. This $(1+z)$ rescaling factor can be safely applied provided that the redshift of collapse is $\gtrsim 2$, i.e. for halo masses below $\sim10^{12}\msunh$: at later times, the effect of the cosmological constant may make advisable the use of a more accurate scaling factor, as discussed e.g. in M08. The concentration of an individual dwarf-size halo with a mass of $\sim$10$^8\msunh$ has also been included in the top panel of Fig.~\ref{fig:cm0} (green pentagon) that comes from a pioneering dwarf galaxy high resolution simulation work done by \citet{moore01}. In this case, the original concentration value in \citet{moore01} has been extrapolated from $z=4$ down to the present epoch by means of the $(1+z)$ factor. We also rescued 218 halos between $\sim 10^6 - 10^{9} \msunh$ from the Via Lactea II (VL-II) simulation \citep{VLII}. In this case, we performed a search of those distinct halos that i) are located between $1-1.5$ times the virial radius of the main halo, ii) were not subhalos at earlier times either, attending to their velocity histories\footnote{More precisely, we apply the restriction that the maximum circular velocity reached by the halo over its entire existence is the one at $z=0$ within 5\%.}, and iii) do not belong to the other parent halo that is present in the 
VL-II simulation box. We calculated median halo concentrations and $1\sigma$ errors after having grouped the 218 halos in 5 mass bins (red triangles). We also display in the top panel of Fig.~\ref{fig:cm0} the results reported for Earth-mass microhalos ($\sim$10$^{-6}\msunh$) at $z=26$ and $z=31$ by \citet{diemand05} and \citet{anderhalden13}, filled and empty diamonds respectively, as well as the upper limit to halo concentrations mentioned by \cite{diemand05} in the range $\sim$10$^{-6} -$10$~\msunh$ (pink dotted line). Note that, in the case of the mentioned microhalo concentrations, each data point in Fig.~\ref{fig:cm0} corresponds to an {\it individual} microhalo. It is only very recently tha, \citet{ishiyama14} has been able to measure microhalo concentrations with an excellent statistics between $4\times10^{-4} \msun$ (few dozens of halos) down to $\sim 2\times10^{-6} \msun$ (more than 2,000 halos; T. Ishiyama, private communication). We reproduce in Fig.~\ref{fig:cm0} their median concentrations (plus the associated 25 and 75 per cent quartile values) found at $z=32$ in their {\it A\_N4096L400} simulation box (empty black triangles). All mentioned microhalo concentration values were scaled to the present time by applying the $(1+z)$ correction factor. 

Most of the existing c(M) median relations proposed in the literature were derived from well resolved simulations in the halo mass range from 10$^{10}$ -- 10$^{15}\msunh$, e.g., B01, \citet{hennawi07,neto07,duffy08}, M08, \cite{munozcuartas11}, P12. Usually, these relations have been approximated by power-laws, as they are good fits over the relatively small mass range considered in those works. Yet, we recall that c(M) power-law models are not expected over the full halo mass range shown in the top panel of Fig.~\ref{fig:cm0}, i.e. from massive clusters down to the lightest Earth-like micro-halos, given the non power-law behavior of $\sigma(M)$ in the $\Lambda$CDM cosmology and, ultimately, given the shape of the matter power spectrum. 
There are, however, c(M) models that link concentration with the $r.m.s.$ amplitude of the linear density field fluctuations and are thus better physically motivated. 
We show in Fig.~\ref{fig:cm0} the model recently proposed by P12, together with the refined version of the B01 model provided by M08 (solid and dashed black lines, respectively). For the latter, we used the parameters derived in M08 for {\it all} halos in the WMAP5 cosmology. In Fig.~\ref{fig:cm0}, we also display a typical concentration scatter of 0.14 dex centered on the P12 model (shaded blue region) following previous estimates, e.g., B01;~\citet{wechler02,dolag04};~M08. In particular, by comparing the simulation data, including the low-mass regime, against the P12 model, we realize that the agreement is remarkably good, with the model being compatible with the measured concentrations within about $1\sigma$.  
Note the excellent agreement between the P12 model and data despite the fact that the mass range spans over 22 orders of magnitude and the slightly variations in the cosmological parameters used in every simulation. This is the case of the $P(k)$ normalization $\sigma_8$, i.e., $\sigma(R=8~\mpc)$, which adopted for the different simulation data set values in the range $0.7<\sigma_8<0.9$, and the matter density, $\Omega_m$, which varies between $0.238<\Omega_m<0.3$.\footnote{We have corrected the virial masses and concentrations for each simulation data set to the 200$\rho_{crit}$ overdensity definition adopted in P12.} The P12 model, which uses $\sigma_8=0.82$ and $\Omega_m=0.27$, works remarkably well even at the smallest 10$^{-6}\msunh$ halo masses.

The P12 model was derived from the Bolshoi and Multidark simulation data. Therefore, one may ask whether the extrapolation of the model down to the minimum halo mass in the top panel of Fig.~\ref{fig:cm0} (and also recently shown in \citet{NG}) is entirely justified, and thus if the agreement between P12 and low-mass concentration data is accidental or not. The fact is that, indeed, there is not extrapolation. The latter can be better understood by plotting the halo concentration against $\log[\sigma(M)]^{-1}$, after having rescaled the results to $z=0$ in a particular way (bottom panel of Fig.~\ref{fig:cm0}). Following P12, this rescaling only affects $\sigma(M)$: halo concentration values are those in the original set of simulations at the corresponding redshift where they were measured, while $\sigma(M)$ values are the ones that halos would have at present time for {\it those} values of the concentration.\footnote{More precisely, and following the nomenclature in P12, we plot $\mathcal{C}(\sigma')$, i.e., their Eq.~(16).} In such $c$~--~$\sigma(M)^{-1}$ plane, the P12 model adopts a characteristic U-shape, with its minimum value corresponding to the natal concentration of DM halos at $z=0$. We propose that halo evolution tracks follow this U-shape from right to left, in such a way that halos found to the right of the minimum ($\sigma < 1$) are not formed yet, while halos located to the left have already collapsed. This is supported by the fact that at the high-mass end ($\sigma < 1$) the median halo kinematic profiles show large signatures of infall and highly radial orbits (see P12). As the P12 model was derived and tested between $-0.5 \lesssim \log[\sigma(M)]^{-1} \lesssim 0.5$ (i.e., the range around the U-shape minimum) by using Bolshoi and Multidark data at different redshifts, the model can be safely used to predict median concentration values of any simulation data whose $\sigma(M)$ values lie within that particular tested interval of the U-shape. As shown in the bottom panel of Fig.~\ref{fig:cm0}, this is exactly the case for almost all the c(M) data set displayed in the top panel of the same figure. Thus, no extrapolation of the P12 model is done at the smallest scales, which also explains its remarkable agreement with the data in that lower mass regime. 
We note that, for the bottom panel of Fig.~\ref{fig:cm0}, we are actually correcting by the different $\sigma_8$ values used in some of the simulations by taking the right $P(k)$ in each case, namely MulitDark, Bolshoi, and the dwarfs simulations by \citet{colin}.

Finally, we provide a simple parametrization of the median concentration-mass relation provided by the P12 model at $z=0$, that will turn out to be very useful for the next section, where we will compute the expected substructure halo boosts to the DM annihilation signal: 

\begin{equation}
    c_{200}(M_{200},z=0) = \sum_{i=0}^{5} c_i \times \left[ \ln \left( \frac{M_{200}}{\msunh} \right) \right]^i ,
    \label{eq:param}
\end{equation}
           
\noindent where $c_i = [37.5153, -1.5093,$ $1.636\cdot 10^{-2}, 3.66 \cdot 10^{-4}, $ $-2.89237\cdot 10^{-5}, 5.32 \cdot 10^{-7}]$. This parametrization, inspired from the functional form proposed by \citet{lavalle08}, provides an accuracy better than 1\% in the halo mass range between $10^{-6}  < M_{200}~\msunh < 10^{15}$. It also captures the characteristic c(M) upturn at higher masses found in P12.\footnote{This upturn seems to be caused by the lack of full relaxation in a large fraction of massive haloes, see e.g. \citet{ludlow12}.} We note that, interestingly, the best fit to VL-II  (subhalo) concentrations found by \citet{pieri11} agrees very well with Eq.(\ref{eq:param}) in the mass range well resolved in that simulation, i.e. $10^{5}  \lesssim \msunh \lesssim 10^{9}$, with deviations becoming only relevant at lower and, very specially, higher masses.

\section{Halo substructure boosts to the dark matter annihilation signal} \label{sec:boosts}
An important open question today is the role of DM substructure in $\gamma$-ray DM searches. Indeed, DM substructure might represent the key component in future DM search strategies for several reasons. In particular, as the DM annihilation $\gamma$-ray signal is proportional to the DM density squared, the clumpy distribution of subhalos inside larger halos expected in $\Lambda$CDM may boost the DM annihilation flux considerably. This flux enhancement is more important for the most massive halos as they enclose more hierarchical levels of structure formation. The effect of substructures on the DM annihilation flux (frequently known as {\it substructure boost}) has already been studied extensively both analytically, e.g., \citet{pieri08b,lavalle08,martinez09}, and making use of $N$-body simulations, e.g., \citet{kuhlen08,springel}. 
It is a challenge to estimate analytically the survival probabilities of substructures within their host halos, while state-of-the-art $N$-body simulations are computational prohibited to simulate the sub-halo hierarchy below a mass $\sim$10$^5 \msunh$, being still very far from the predicted halo cut-off mass, of the order of 10$^{-6}\msunh$ or even smaller, e.g., \citet{green04,minmass}. 

Most popular substructure boost models (e.g.,~\citet{pinzke11,gao11}) implicitly rely on power-law extrapolations of the c(M) relation well below the resolution limit of $N$-body simulations all the way down to the minimum halo mass. Thus, any power-law extrapolations will assign very high concentrations to the smallest halos. As the annihilation luminosity of a given halo scales as $L \propto c^3$, the substructure boosts obtained in this way are usually very large. Furthermore, the results are very sensitive to the power-law index used in such extrapolations. However, as mentioned above, these power-law extrapolations are not expected in the CDM cosmology. Indeed, as small halos over a broad range of masses collapse at nearly the same time in the early Universe, and natal concentrations are set by the halo formation epoch, we expect that low-mass halos have rather similar natal concentrations, and thus will also possess similar concentrations at the present time. This fact translates in a flattening of c(M) at low masses, which is evident in the top panel of Fig.~\ref{fig:cm0} as described by the P12 model, and highlighted by the available low-mass halo data also shown in this figure. We remark that, ultimately, natal halo concentrations are the key for this to happen. 

In the following, we will compute the substructure boosts implied by the P12 model. We note that by doing so we assume this model to be also a good representation of the subhalo concentrations. 
Concentrations of field halos should be a fair estimate of those typical of subhalos of similar mass. Nevertheless, subhalos are known to have slightly higher concentrations, the closer they lie from their host halo centers the larger are their concentrations, e.g.,~\citet{diemand07,pieri11}. On the other hand, sub-subhalo abundance may be reduced compared to subhalo abundance. Thus, overall, the P12 substructure boosts will represent a fair estimate of their actual values. 

\begin{figure*}
\includegraphics*[scale=0.43]{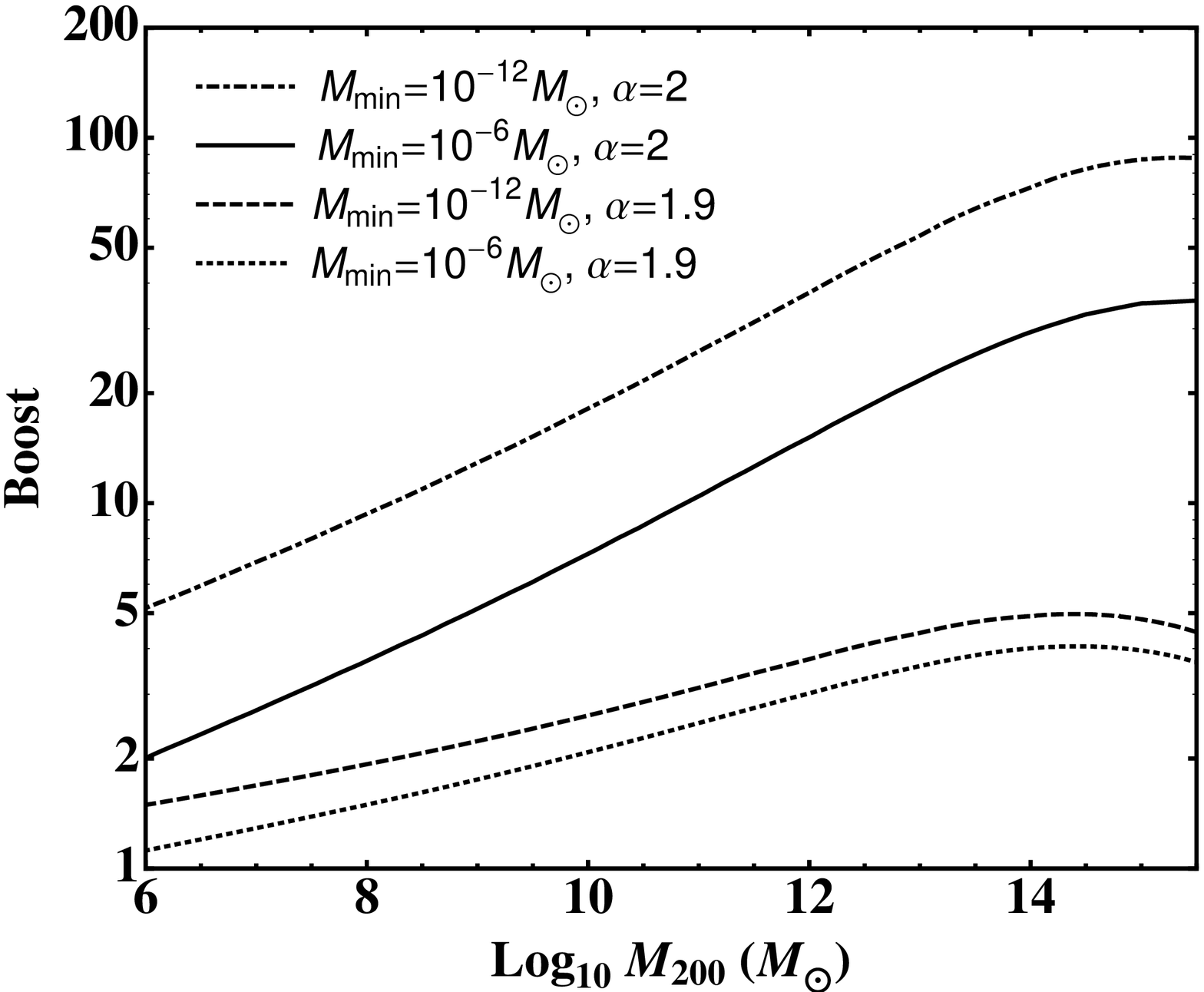}
\includegraphics*[scale=0.43]{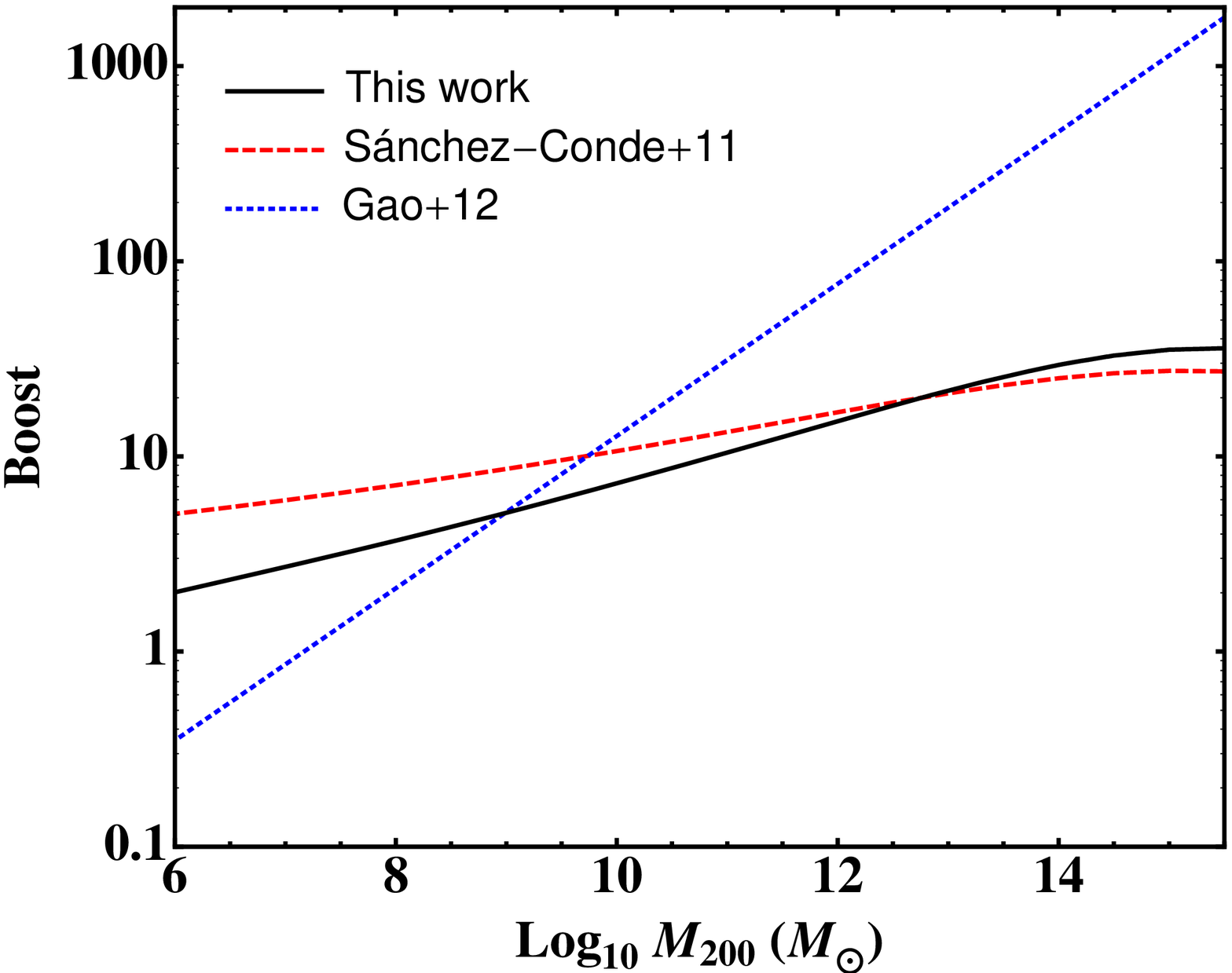}
\caption{{\it Left panel:} Halo substructure boosts as a function of host halo mass obtained with the P12 c(M) model, for different values of minimum subhalo mass, $M_{min}$, and slope of the subhalo mass function, $\alpha$. From bottom to top, the different lines correspond to (M$_{min}$, $\alpha$) = ($10^{-6} M_\odot$, 1.9), ($10^{-12} M_\odot$, 1.9), ($10^{-6} M_\odot$, 2), ($10^{-12} M_\odot$, 2). The solid line corresponds to our fiducial boost model, i.e., $M_{min}=$10$^{-6}\msun$ and $\alpha=2$.  {\it Right panel:} Comparison between the substructure boosts given by our fiducial boost model (solid line), and that computed by \citet{masc11} and \citet{gao11} (dashed and dotted lines, respectively).}
\label{fig:boosts}
\end{figure*}

To compute the boosted annihilation luminosity of a given halo of mass $M$ due to its substructure population, it is necessary to integrate the subhalo annihilation luminosities all the way down to the minimum subhalo mass, $M_{min}$. Since subhalos also host sub-substructure, ideally, all levels of substructure should be included in our boost calculation. We define the boost $B(M)$ as follows \citep{strigari07,kuhlen08}:

\begin{equation}
B(M) = \frac{1}{L(M)}\int_{M_{min}}^M (dN/dm)~[1+B(m)]~L(m)~dm,
    \label{eq:boost}
\end{equation}

\noindent where $L(M) = 4\pi Mc^3/f(c)^2$ is the halo annihilation luminosity with no substructures, $c$ being the concentration, $f(c)=\log(1+c)-1/(1+c)$, and $dN/dm=A/M~(m/M)^{-\alpha}$ is the subhalo mass function. Values of $\alpha$ ranging between $\alpha=1.9$ and $2$ are possible \citep{diemand07,madau08,springel}. The normalization factor $A$ is chosen to match the amount of substructure resolved in current simulations (typically $\sim$10\%, e.g., \citet{kuhlen08,springel}). We find $A$ to be equal to 0.030 and 0.012 for $\alpha=1.9$ and $2$, respectively. Note that following the definition of boost in Eq.~(\ref{eq:boost}), an scenario with no boosted signal would imply $B=0$, while a value of $B=1$ would mean that substructures contribute to the annihilation luminosity at the same level than the parent halo. We show in Fig.~\ref{fig:boosts} the results of computing the substructure boost with Eq.(\ref{eq:boost}) and using the c(M) parametrization given in Eq.(\ref{eq:param}) for the P12 model. We adopt $M_{min}=10^{-6} M_{\odot}$ and $\alpha=2$ for our fiducial substructure boost model\footnote{The choice of $\alpha=2$ for our fiducial model is motivated by theoretical expectations in the Press-Schechter theory for structure formation, see e.g. \citet{giocoli08,blanchet12}.}, but we also show the result of varying these parameters in the left panel of Fig.~\ref{fig:boosts}. In our computation of the substructure boosts, only the first two levels of substructure were included, i.e., subhalos and sub-subhalos, since according to our checks the third substructure level contributes only less than 5\% to the total boost in most cases (reaching $\sim$8\% in the most extreme case adopting $M_{min}=10^{-12} M_{\odot}$ and $\alpha=2$). The marginal relevance of level 3 was already pointed out by \citet{martinez09}, who analytically predicted a $\sim$2\% signal increase from level 3 and beyond. We note that we find slightly higher contributions from this level though. Level 2, however, can contribute up to one third of the boost value given in our fiducial model for the largest halo masses.

The right panel compares our fiducial boosts with those previously derived by \citet{masc11} and \citet{gao11}. As it can be clearly seen, the boosts yielded by the P12 model qualitatively agree with previous estimates that also used physically motivated c(M) models well below the mass resolution limits of $N$-body cosmological simulations \citep{lavalle08,kuhlen08,pieri08b,martinez09,3k10,charbonnier11,masc11,kuhlen12,nezri12,anderhalden13,zavala13}. These are, however, in clear contradiction with that found in works that implicitly adopted a power-law c(M) extrapolation to lower masses, e.g.,~\citet{springelnature,zavala10,pinzke11,gao11}. For Milky Way-size halos, our fiducial substructure boost model yields a boost of $\sim$15 versus $\sim$77 in the model by \citet{gao11}. The difference is even more pronounced for larger halos, as expected. For a rich 10$^{15}\msun$ galaxy cluster, for instance, we obtain a boost of $\sim$35, while \citet{gao11} estimated $\sim$1100, i.e. about 1.5 orders of magnitude larger! This disagreement would have been even larger if we had compared both approaches for $M_{min}=10^{-12} M_{\odot}$ instead of 10$^{-6} M_{\odot}$: our boosts do not change drastically by including smaller substructures, while power-law-based substructure models are very sensitive to the adopted value of $M_{min}$. On the other hand, note that we do expect a substantial flux increase of a factor of a few due to DM substructure in dwarf galaxies. We recall, however, that strictly speaking our results are only applicable to field halos; for the dwarf galaxies satellites of the Milky Way, for example, tidal stripping may have removed most of the substructure in the outer regions of these objects -- where subhalos typically reside -- in this way significantly decreasing this substructure boost value.\footnote{Actually, sub-subhalo abundance is found to be reduced considerably compared to subhalo abundance (at a fixed mass), see e.g. Figs. (16) and (17) in \citet{springel}.} This decrease may be compensated though by the fact that subhalos are known to exhibit larger concentrations compared to that of field halos \citep{VLII}. We conclude that the final boost value for these objects is not clear at the moment and should be addressed in future work, our results in Fig.~\ref{fig:boosts} representing a first order estimate.

Finally, we provide a simple parametrization for the substructure boost factors implied by the P12 concentrations at $z=0$ for our fiducial model with $M_{min}=10^{-6} M_{\odot}$ and $\alpha=2$ (solid lines in both panels of Fig.~\ref{fig:boosts}), i.e.:

\begin{equation}
    \log_{10} B(M_{200}, z=0) = \sum_{i=0}^{5} b_i \times \left[ \ln \left( \frac{M_{200}}{M_{\odot}} \right) \right]^{i},
    \label{eq:boostsparam}
\end{equation}
        
where $b_i = [-0.442, 0.0796, -0.0025, 4.77\cdot 10^{-6},$ $4.77\cdot 10^{-6}, -9.69\cdot 10^{-8}]$. The accuracy of this parametrization is better than 5\% in the mass range $10^{6}  < M_{200}~\msun < 10^{16}$. We recall that a value of $h=0.7$ was implicitly adopted in both Fig.~\ref{fig:boosts} and Eq.~(\ref{eq:boostsparam}).



\section{Summary} \label{sec:summary}
In this work, we have examined the concentration-mass relation of CDM halos at present time ranging from Earth-mass microhalos up to galaxy clusters. Our current knowledge of the median c(M) relation is summarized in the top panel of Fig.~\ref{fig:cm0}, which shows concentrations  for the entire halo mass range available in the literature from state-of-the-art $N$-body cosmological simulations. A comparison between both, cosmological simulations and c(M) models has allowed us to extract meaningful conclusions. We showed that only realistic models that link halo concentration with the amplitude of the linear density field fluctuations, $\sigma(M)$, such as the 
recent model provided by \citet{prada12}, are a good representation of what is measured in simulations at all scales. Both simulation results and these physically motivated c(M) models show a clear flattening of c(M) at lower masses that excludes the use of simplistic power-law extrapolations down to the smallest scales: this behavior is neither expected in the current CDM cosmological paradigm nor supported by simulations. We also provided a simple parametrization for the c(M) relation at $z=0$, based on the P12 model, that spans over 22 orders of magnitude in halo mass.

Since halo annihilation luminosity is a strong function of the concentration, the flattening of c(M) at the smallest scales is expected to have a large impact on gamma-ray DM search studies. As a particular example, we used the c(M) model by \citet{prada12} to compute the so-called substructure boosts factors, and found much more modest boosts than those recently discussed in the literature, e.g., \citet{springelnature,pinzke11,gao11}. These works implicitly adopted simple power-law concentration models down to the minimum halo mass, which led to a wrong overestimation of halo concentrations at the smallest scales and therefore of the boost. Furthermore, their results are extremely sensitive to the exact choice of the minimum halo mass, contrary to what happens in the case of using physically motivated c(M) models. We provided a parametrization of the substructure boost assuming the \citet{prada12} for the subhalo concentrations. This parametrization works remarkably well for objects in the mass range between that of dwarfs and clusters. We recall that the substructure boosts that we found rely on a c(M) model that is strictly valid only for field halos, and it is known that, in comparison, subhalos exhibit higher concentrations. This may lead to higher boosts. On the other hand, loss of material in the outskirts of subhalos due to tidal interactions may reduce the contribution of lower levels of substructure in these objects, leading to lower boost values. We conclude that the actual boost values will depend on the exact interplay between the mentioned effects, our boosts representing a fair first order estimate. A more refined substructure boost model that account for these and other potential effects is left for future work.

Our work has been particularly useful to highlight where we stand on understanding the nature of halo concentrations and to also identifying new opportunities for further theoretical and numerical investigation. For instance, we identified a clear absence of simulation data below $\sim10^8 \msunh$ which should be ideally covered and studied by new high resolution $N$-body simulations. The challenge is twofold: we want to simulate not only the smallest mass scales -- which have very specific and inherent difficulties -- but also to have a good halo statistics that allow us to extract robust and statistically meaningful results. If succeeded, CDM halo concentrations model predictions will be tested up to a higher degree of accuracy, enabling new avenues of both studying the internal properties of CDM halos over a huge range of halo masses, and of testing this particular cosmological structure formation framework.

\section{Acknowledgements}
We thank J\"urg Diemand, Ra\'ul Angulo, Donnino Anderhalden, Manoj Kaplinghat, Anatoly Klypin, Aaron Ludlow and Joel Primack for enlightening discussions during the completion of this work. We also thank Pedro Col\'in, Ginevra Favole and Tomoaki Ishiyama for providing with crucial details of the simulation data in Fig.~\ref{fig:cm0}. In particular, we are also grateful to R. Angulo, J. Diemand and J. Primack for useful comments on this manuscript. The work of MASC is supported through the NASA grant NNH09ZDA001N. The authors also acknowledge support from the MULTIDARK project of Spanish MCINN Consolider-Ingenio: CSD2009-00064. MASC is also grateful for the hospitality of the Instituto de F\'isica Te\'orica (IFT-UAM/CSIC) in Madrid, where part of this work was done, and the continuous and invaluable help of Susana Hern\'andez on logistics issues.

\end{document}